\documentclass[%
 reprint,
 amsmath,amssymb,
 aps,
]{revtex4-2}

\usepackage{graphicx}
\usepackage{dcolumn}
\usepackage{bm}
\usepackage{amsmath}

\begin{document}

\preprint{APS/123-QED}

\title{Parallel spatial photonic Ising machine using spatial multiplexing for accelerating combinatorial optimization}

\author{Suguru Shimomura}
 \email{s-shimomura@ist.osaka-u.ac.jp}
 \author{Jun Tanida}%
 \author{Yusuke Ogura}%
\affiliation{Graduate school of Information Science and Technology, Osaka University.}%

\date{\today}

\begin{abstract}
A spatial photonic Ising machine (SPIM) handles large-scale combinatorial optimization problems owing to optical processing with spatial parallelism. However, iterative  feedback in the search for optimal solutions limits processing speed even though the Ising Hamiltonian is computed optically. 
We propose a parallel spatial photonic Ising machine (pSPIM) utilizing spatial multiplexing to search for an optimal solution efficiently. 
By employing grating patterns and encoding multiple sets of Ising spins in a phase distribution, several Ising Hamiltonians are computed simultaneously. 
We demonstrated that Max-Cut problems requiring 100 Ising spins are solved faster as the number of processing units increases. 
In addition, combining the multicomponent model with parallel processing allows for efficient searching for optimal solutions to problems represented by using interaction matrixes with a rank greater than one. 
The pSPIM achieves high-speed searching of optimal solutions of large-scale combinatorial optimization problems.

\end{abstract}

\maketitle


{\it Introduction.}---Modeling physical phenomena contributes to solve problems in our information society, in applications handling large amounts of data.
An Ising machine, which is implemented based on a model of magnetic behavior, can solve combinatorial optimization problems that are categorized as nondeterministic polynomial hard \cite{Barahona1982,Lucas2014,Mohseni2022}. 
Iterative sampling based on heuristic algorithms and energy minimization in the Ising model formulating the problem allows searching for optimal solutions. 
Various Ising machines have been proposed to search for solutions by employing optical and electronic parametric oscillators \cite{Inagaki2016,Honjo2021,Bohm2019,Cen2022,Marandi2014}, CMOS devices \cite{Yamaoka2016}, nanomagnets \cite{Sutton2017}, and superconducting qubits \cite{Albash2018,Farhi2001,Johnson2011,Kadowaki1998,Santoro2006}.
A spatial photonic Ising machine (SPIM) is a promising system for solving large-scale combinatorial optimization problems by encoding Ising spins into phase distributions \cite{Pierangeli2019}. 
Compared to other Ising machine architectures, the SPIM optically computes the Ising Hamiltonian with all-to-all interactions between more than $1 \times 10^4$ Ising spins \cite{Prabhakar2023}.
Moreover, the computation time required to obtain the Ising Hamiltonian is constant regardless of the number of the Ising spins.
To utilize spin scalability and address various combinatorial optimization problems, many approaches have been proposed to enhance the applicability of the SPIM, for example, by improving the annealing algorithm \cite{Pierangeli2020,Ramesh2021}, by developing novel spin encoding methods \cite{Sun2022,Sakellariou2024,Ye2023,Wang2024efficient,Ouyang2024}, and by utilizing physical phenomena to represent interaction matrices \cite{Kumar2023,Pierangeli2021}. Furthermore, multicomponent and eigen decomposition methods using optical techniques including time \cite{Yamashita2023}, space \cite{Sakabe2023}, and wavelength division multiplexing \cite{Luo2023} expand the degrees of freedom to represent interaction matrices, which is necessary to address a wide range of problems. 

Previous methods have focused on solving problems with multiple constraints. However, the computing speed to find the optimal solution is inherently slow. Switching via electronic feedback to a spatial light modulator (SLM) that encodes the Ising spins into the phase distribution regulates the processing speed of the SPIM. As the number of Ising spins and the constraints increase, solution candidates and the problem complexities also increase. To find the optimal solution, an extremely long annealing time is required, and it becomes unavoidable to switch a screen of the SLM enormous number of times. As a result, extensive computation times is required to optimize the set of Ising spins. 

In this letter, we propose a parallel SPIM (pSPIM) to search for optimal solutions effectively and to avoid the enormous iterations of electronic feedback. By employing a spatial multiplexing technique, simultaneous computation of multiple components of the Ising Hamiltonian can be implemented \cite{Sakabe2023,Shimomura2023,Veraldi2025}, and therefore, updating the Ising spins based on multiple Ising Hamiltonians reduces the number of electronic feedback cycles.
We evaluate the performance of the proposed system in searching for optimal solutions to Max-Cut and knapsack problems and assess the effectiveness of the parallel processing approach.

\begin{figure}[bt]
  \centering
  \includegraphics[width = 7.5cm]{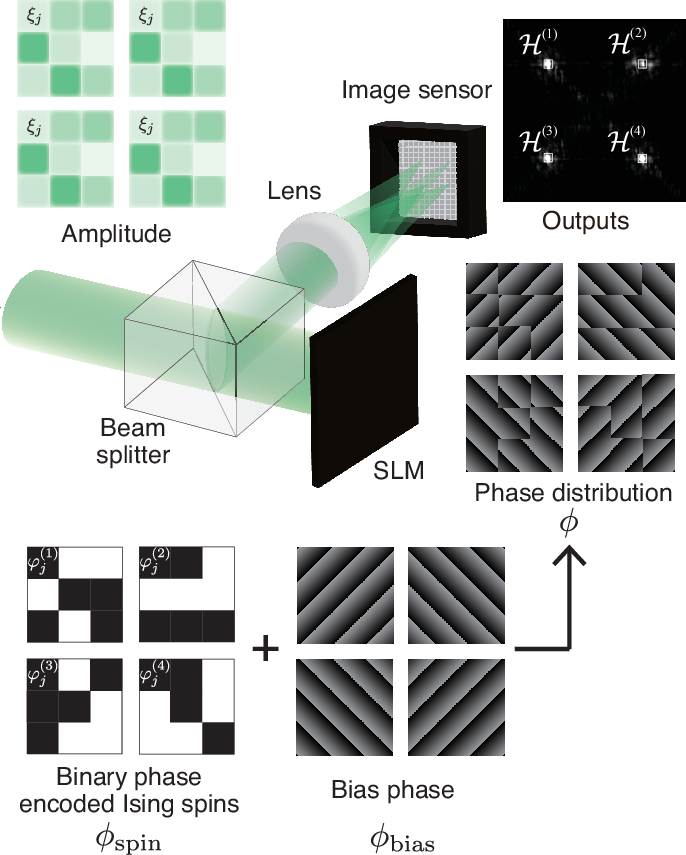}
  \caption{Schematic of pSPIM. Light patterns with amplitude distributions $\xi^{(k)}$ is radiated onto an SLM which encodes multiple Ising spins $\sigma^{(k)}$ into phase distribution $\phi_{\rm spin}^{(k)}$. The light intensities representing Ising Hamiltonians corresponding to individual spin states are obtained by an image sensor.}
  \label{fig:concept}
\end{figure}

{\it Parallel processing  by using spatial multiplexing.}---The SPIM computes the Ising Hamiltonian by encoding $N$ Ising spins and their interactions into the phase and amplitude of light, respectively. 
The Ising spin $\sigma_j=\exp{\{i\phi_{j}\}}\in \{-1,1\}$ with $j=1,\cdots,N$ is encoded into the binary phase $\phi_j$ of the light with amplitude $\xi_j$. By using an SLM, the light distribution is set to:
\begin{equation}
E (s,t) = \sum_j \xi_j\sigma_j \tilde{\delta}_W(s-2Ws_j)\tilde{\delta}_W(t-2Wt_j), 
\end{equation}
where $s$ and $t$ represent two dimensional coordinates at the SLM plane. $\tilde{\delta}_W$ is the normalized rectangular function representing the macropixel area of the SLM with size $2W$, and $(s_j, t_j)$ corresponds $j$-th macropixel. After passing through a lens, the light intensity distribution at the focal plane is given by
\begin{multline}
    I(x,y) = \sum_{j,h} \xi_j\xi_h\sigma_j\sigma_h\delta^2_W(x)\delta^2_W(y)\\
         \times e^{2iW(\{s_j-s_h\}x+ \{t_j-t_h\}y)},
  \end{multline} 
where $x$ and $y$ represent the two-dimensional coordinates on the focal plane, and $\delta_W(x) = {\rm sinc}(Wx)$ is the inverse Fourier transformation of $\tilde{\delta}_W$. At the focal point, $x=y=0$, the intensity value corresponds to the Ising Hamiltonian $\mathcal{H}=I(0,0)= \sum_{j,h} \xi_j\xi_h\sigma_j\sigma_h$ with spin-spin interaction $J_{jh}=\xi_j\xi_h$. An optimal solution can be sought by minimizing the Ising Hamiltonian based on an update of the binary phase distribution representing a set of Ising spins. 
To implement parallel processing by multiplexing operation, multiple amplitude distributions are arranged spatially. Figure \ref{fig:concept} shows a schematic diagram of the pSPIM. For parallel processing of the Ising Hamiltonians,
the amplitude pattern $\xi = (\xi_1,\xi_2,\cdots,\xi_N)$ is replicated in $K$ multiplexing operation, and bias phase distribution $\phi_{\rm bias}^{(k)}$ with a grating pattern is added to the phase distribution $\phi^{(k)}_{\rm spin} = \sum_j \phi_j^{(k)} \tilde{\delta}_W(s-2Ws_j)\tilde{\delta}_W(t-2Wt_j)$ with $k=1,\cdots, K$:
\begin{align}
  &\phi_{\rm bias}^{(k)} = \alpha^{(k)}s+\beta^{(k)}t\\
  &\phi^{(k)} = \phi^{(k)}_{\rm spin} + \phi_{\rm bias}^{(k)},
\end{align}
where $\alpha^{(k)}$ and $\beta^{(k)}$ represent coefficients corresponding to diffraction angles of the grating pattern in $k$-th multiplexing operation. The phase distribution of individual light pattern is set to $\phi^{(k)}$ by using the SLM.
The captured intensity distributions in the $k$-th multiplexing operation on an image sensor are described by:
\begin{multline}
  I^{(k)}(x,y) = \sum_{j,h} \xi_j\xi_h\sigma_j^{(k)}\sigma_h^{(k)}  \delta_W^2(x+{\alpha}^{(k)})\delta_W^2(y+\beta^{(k)}) \\ 
  \times e^{2iW\{(s_j-s_h)(x+\alpha^{(k)})+(t_j-t_h)(y+\beta^{(k)})\}}.
\end{multline}
The intensity value at position $(x,y) = (-\alpha^{(k)}, -\beta^{(k)})$ corresponds to the Ising Hamiltonian $\mathcal{H}^{(k)}=\sum_{j,h}\xi_j\xi_h\sigma_j^{(k)}\sigma_h^{(k)} $ for the set of the Ising spin $\sigma^{(k)} = (\sigma_1^{(k)},\sigma_2^{(k)},\cdots,\sigma_N^{(k)})$. Owing to the two-dimensional direction of the grating patterns, individual Ising Hamiltonians are arranged separately. As a result, multiple Ising Hamiltonians are obtained simultaneously without switching the pattern displayed on the SLM. 
Moreover, updating of the Ising spins from multiple Ising Hamiltonian contributes to the search for better solution efficiently and shortens the annealing time  to find the optimal solution in simulated annealing \cite{Ram1996}. 
Therefore, the pSPIM accelerates the commutation time not only by reducing the number of electronic feedbacks but also by searching for optimal solutions efficiently. 
Furthermore, parallel processing can be applied for computing the Ising Hamiltonian representing the problems with multiple constraints. The Ising Hamiltonian can be decomposed into several components and computed by linear combination of the outputs obtained from the SPIM \cite{Yamashita2023,Sakabe2023,Luo2023}. By assigning different amplitude distributions $\xi^{(k)} $ and the same phase distribution $\phi$ to each multiplexing units, the pSPIM can handle a combinatorial optimization problem with a large number of constraints.

{\it Experimental setup and algorithm to update Ising spins.}---Light patten with amplitude distribution $\xi^{(k)}$ according to $K$ multiplexing units is generated by using a liquid-crystal-on-silicon (LCOS) SLM (Santec, SLM-200, 1920$\times$1200 pixels with pixel pitch of 8.0 $\mu$m) and laser beam emitted from a diode-pumped solid-state laser (Shanghai Sanctity Laser, SSL-532-0050-LN) with a wavelength of $\lambda =$ 532 nm. 
The generated pattern is radiated onto a phase-modulation LCOS SLM (Hamamatsu Photonics, X15213-01, 1272$\times$1024 pixels with a pixel pitch of 12.5 $\mu$m), and the phase distributions are set to $\phi^{(k)}$ in the $k$-th multiplexing operation. 
The modulated pattern is focused through a lens (focal length $f = 300$~mm) and multiple Ising Hamiltonians are obtained from the captured image by a CMOS image sensor (Point Grey Research, GS3-U3-51S5M). 
Among the obtained Ising Hamiltonians, the lowest value is selected, and the best of the Ising spins is updated after comparing it with the previously selected Ising Hamiltonian using the Metropolis algorithm \cite{Van1987}. 
The Ising spins in the next annealing time at each multiplexing unit are sampled by a Markov-chain Monte Carlo method from a Gibbs distribution.
To obtain the Ising Hamiltonian $H^{(k)}$, it is necessary to determine individual points $(x,y)= (-\alpha^{(k)},-\beta^{(k)})$. 
For this purpose, only bias patterns with $\phi_{\rm bias}^{(k)}$ are displayed on the SLM before the computation, and the maximum-intensity pixel on the image sensor are identified as $(x,y)= (-\alpha^{(k)},-\beta^{(k)})$.

{\it Experimental result in searching for optimal solution}---.
\begin{figure*}
  \centering
  \includegraphics[width = \linewidth]{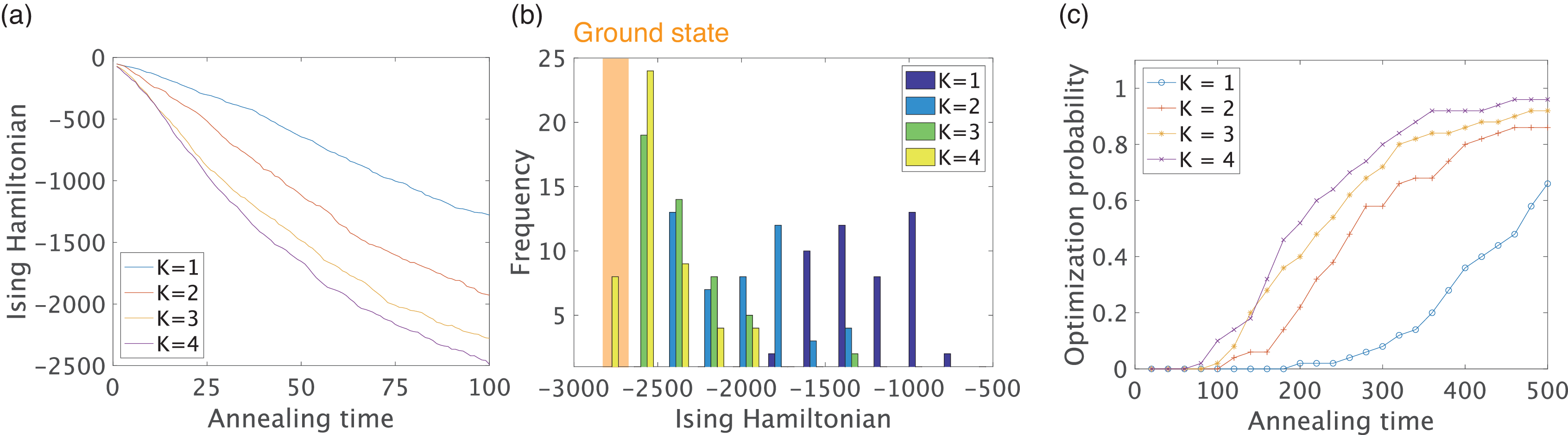}
  \caption{(a) Time evolution of averaged Ising Hamiltonian depending on $K$ multiplexing units in solving the Max-Cut problem. (b) Histograms of the energy value of 50 samples for $K$ multiplexing units. (c) Optimization probabilities depending on annealing time.}
  \label{fig:maxcut_time_hist}
\end{figure*}
To evaluate the performance of the pSPIM, an optimal solution to the Max-Cut problem for $N= 100$ was sought. The Max-Cut problem is a benchmark task to evaluate searching capabilities for optimal solutions.
In this study, the interaction matrix was randomly set to $J_{jh}\in \{-1,0\}$. The macropixel size $W$ was set to $200 {~\rm \mu m}$, which is the least common multiple of the pixel sizes of the amplitude and phase modulation SLM. Solutions were obtained over 100 iterations under constant temperature, and the best set of the Ising spins indicting the solutions was recorded during the annealing.

Figure \ref{fig:maxcut_time_hist}(a) shows annealing time evolution of the averaged Ising Hamiltonian over 50 trials.
As the multiplexing units for parallel processing increased, the Ising Hamiltonian values decreased rapidly. The median values of the Ising Hamiltonian after annealing were $-1.2\times10^{-3}, 1.9\times10^{-3}, -2.3\times10^{-3}$ and $-2.5\times10^{-3}$ for each number of multiplexing units.
Histograms of Ising energy after the annealing are shown in Figure \ref{fig:maxcut_time_hist}(b).
In the case of 4 multiplexing units ($K=4$), the ground state of the Ising Hamiltonian was found 8 times in 50 trials, whereas it could not be found in the case of single units ($K=1$). These results show that the pSPIM enables efficient searching for optimal solutions. 
To compare processing performance at the individual numbers of multiplexing units,
we evaluated the annealing time required to achieve an optimal solution. 
The optimization probability was defined as  the ratio of frequency of optimal solutions for 50 trials, and was estimated at each annealing time. The relationship between annealing time and optimization probabilities is shown in Fig. \ref{fig:maxcut_time_hist}(c).
We evaluated the time required to achieve the optimization probability of 0.5. 
The satisfied times for individual multiplexing units were 445, 245, 210, and 164, respectively. 
These results indicate that the computation time with 4 multiplexing units was reduced by a factor of 2.7 compared to a single unit. Increasing the number of multiplexing units contributes to more efficient searching for optimal solutions.

\begin{figure}[bt]
  \centering
  \includegraphics[width = 6.5 cm]{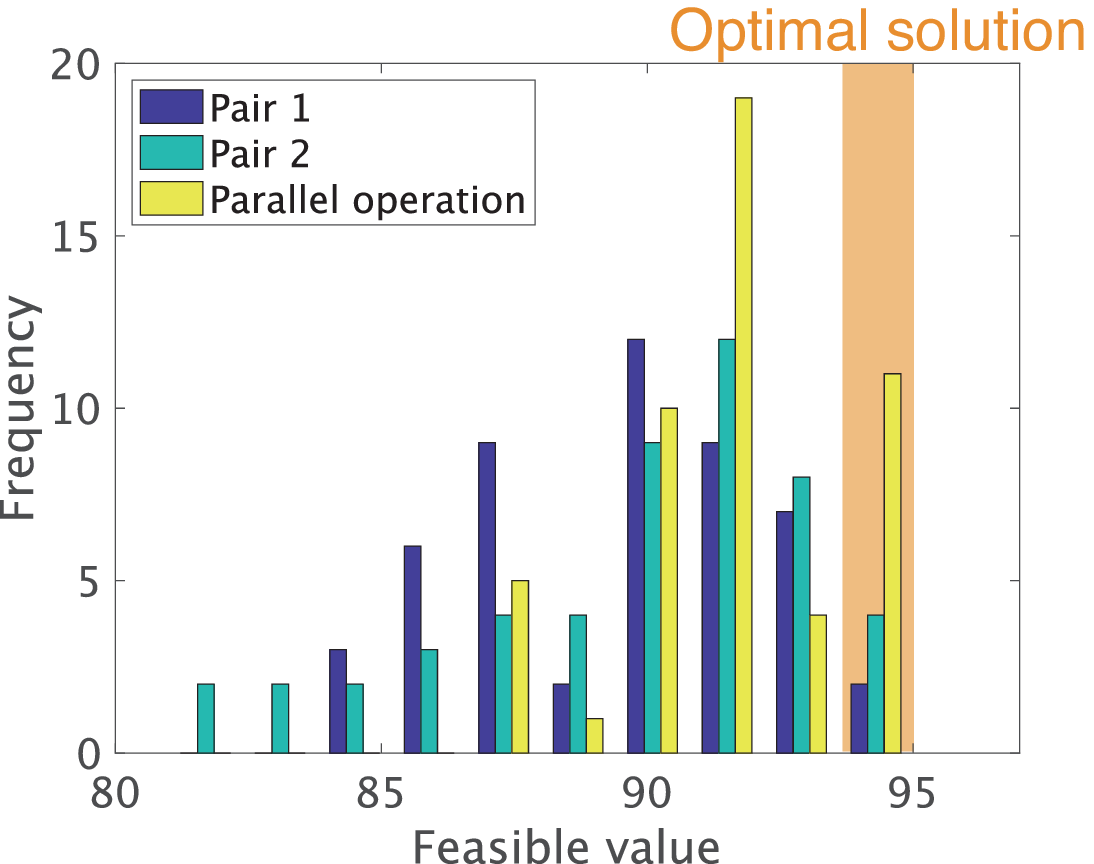}
  \caption{Histograms of feasible solutions in knapsack problems. Pair 1 and 2 indicate a pair of different two multiplexing units, respectively. Histogram of parallel operation is obtained by parallel processing using both pairs.}
  \label{fig:hist_knapsack}
\end{figure}

Next, we applied the pSPIM to solve the knapsack problem, which is a benchmark task for finding an optimal solution that satisfies predefined constraints. 
The Ising Hamiltonian of the 0-1 knapsack problem with integer weights is formulated using the logarithm trick \cite{Lucas2014}:
\begin{align}
  &\mathcal{H} = \mathcal{H}_A+\mathcal{H}_B,\\
  &\mathcal{H}_A = A\left(\mathcal{W}-\sum^n_{i=1}w_ip_i-\sum^m_{i=1}2^{i-1}q_i\right)^2,\\
  &\mathcal{H}_B = B\left(\sum^n_{i=1}v_ip_i\right)^2,
  \end{align}
where $v_i$ and $w_i$ represent the value and weight of the $i$-th item ($i = 1, \cdots, n$), and $\mathcal{W}$ is a limit of the weight. $A$ and $B$ are coefficients of individual terms. $p_i\in\{0,1\}$ and $q_i\in\{0,1\}$ are auxiliary variables. This Ising Hamiltonian is described in multicomponent form \cite{Yamashita2023},

\begin{align}
  &\mathcal{H} = A\sum_{j,h}^n\xi^{(a)}_{j}\xi^{(a)}_{h}\sigma_{j}\sigma_{h}+B\sum_{j,h}^n\xi^{(b)}_{j}\xi^{(b)}_{h}\sigma_{j}\sigma_{h},\\
&\xi^{(a)} = \left(w_1,\cdots,w_n,2^0,\cdots,2^{n-1}, C \right),\\
&\xi^{(b)} = \left(v_1,\cdots,w_n,0,\cdots,0,\sum_i v_i\right),\\
&\sigma = \left(2p_1-1,\cdots,2p_n-1,2q_1-1,\cdots, 2q_m-1,1\right),
\end{align}
where $m = \lfloor \log_2(\max~w_i)\rfloor$, and $C = \sum_i w_i +2^m-1+2\mathcal{W}$. The number of spins is set to $N = n+m+1$.
In the pSPIM, the amplitude distributions consisting of $\xi^{(a)}$ and $\xi^{(b)}$ was set to the irradiation patterns, respectively. 
By modulating both of the phase distributions to $\phi^{(k)}$,
the light intensities $I^{(k_a)}$ and $I^{(k_b)}$ corresponding to the components $\mathcal{H}_A$ and $\mathcal{H}_B$ with Ising spin $\sigma^{(k)}$ were obtained simultaneously.
After weighting the components with coefficients $A$ and $B$ by computational processing, the Ising Hamiltonian was obtained. 
For parallel processing, the amplitude patterns with $ \xi^{(a)}$ and ${\xi^{(b)}}$ are replicated spatially. 
The sets of the Ising spins $ \sigma^{(k)}$ depending on the number of parallel operation were replicated into two respectively to compute multiple sets of the two components $\mathcal{H}_A$ and $\mathcal{H}_B$ simultaneously. 
As a proof of concept, the number of items in the knapsack problem was set to $n=13$. The values and weights of each item in the problems were set by referring to previous research \cite{Yamashita2023,Sakabe2023}. The annealing time was set to 2000, and the temperature in the simulated annealing algorithm was kept constant. After annealing, the best combination obtained samples which met the weight constraints was selected as a feasible solution of the problem. 
Figure \ref{fig:hist_knapsack} shows histograms of feasible solutions obtained from 50 trials. 
Among the 4 units, two types of pair (Pair 1 and 2) were employed to compute the Ising Hamiltonian and search for solutions independently. In parallel processing, Pair 1 and 2 were implemented simultaneously (Parallel operation). The pSPIM with the two sets of multiplexing units found optimal solutions more than twice as often as that with the one.
These results indicate that updating Ising spins based on multiple Ising Hamiltonians simultaneously improves the search capability for optimal solutions compared to independent processing.
Notably, the 0-1 knapsack problem is formulated as an Ising problem with a rank-2 interaction matrix, which cannot be handled by the primitive SPIM systems. Therefore, pSPIM not only enhances the efficiency of searching for optimal solutions but also expands the applicability of the SPIM system.
We compared the computation time of the pSPIM and the SPIM using time-division multiplexing \cite{Yamashita2023}. The averaged time per annealing to solve the knapsack problem was $2.4\times 10^{-1}$ seconds with parallel processing, whereas it was $1.3$ seconds with time-division multiplexing. The pSPIM achieved a processing speed more than 5.4 times faster than the SPIM using time-division multiplexing.
This improvement is due to avoiding modulation of light amplitude. 
The fluctuation of amplitude due to actuation delay of the SLM causes computation errors of the Ising Hamiltonian. To reduce the errors, sufficient time to switch the screen of the SLM completely was required. By avoiding the analog modulation of the amplitude by the SLM, high-speed computation of the Ising Hamiltonian was achieved.

{\it Conclusions.}---We demonstrated a parallel processing system using an SPIM with spatial multiplexing.
By employing grating patterns, simultaneous computation of the Ising Hamiltonian was realized.
The results show that updating Ising spins based on multiple Ising Hamiltonians provides an efficient search for optimal solutions. The pSPIM also has potential applicability to various types of combinatorial optimization problems by utilizing a multicomponent model.

This research was supported by JST-ALCA-Next Program (Grant Number JPMJAN23F2) and JSPS KAKENHI (Grant Number 23H04805). We acknowledge T. Kihara, H. Yamashita, S. Shirasaka and H. Suzuki for useful discussions about analysis of the experimental results.

\bibliography{MyCollection}

\end{document}